\documentclass[a4paper]{jpconf}
\usepackage{url}

\usepackage{graphicx}
\usepackage{enumitem}
\begin{document}
\title{GRID Storage Optimization in Transparent and User-Friendly Way for LHCb Datasets}


\author{M Hushchyn$^{1, 2}$, A Ustyuzhanin$^{1, 3}$, P Charpentier$^{4}$ and  C Haen$^{4}$}
\address{$^{1}$Yandex School of Data Analysis, Moscow, Russia}
\address{$^{2}$Moscow Institute of Physics and Technology, Dolgoprudny, Russia}
\address{$^{3}$National Research University Higher School of Economics, Moscow, Russia}
\address{$^{4}$CERN, Geneva, Switzerland}
\ead{mikhail.hushchyn@cern.ch}

\begin{abstract}
The LHCb collaboration is one of the four major experiments at the Large Hadron Collider at CERN. Many petabytes of data are produced by the detectors and Monte-Carlo simulations. The LHCb Grid interware LHCbDIRAC [1] is used to make data available to all collaboration members around the world. The data is replicated to the Grid sites in different locations. However the Grid disk storage is limited and does not allow keeping replicas of each file at all sites. Thus it is essential to optimize number of replicas to achieve a better Grid performance.

In this study, we present a new approach of data replication and distribution strategy based on data popularity prediction. The popularity is performed based on the data access history and metadata, and uses machine learning techniques and time series analysis methods. 
\end{abstract}

\section{Introduction}
Data replication strategy is essential for the grid performance. On the one hand, too small number of replicas puts high load on the network, leads to non-optimal usage of the computing resources and large job wall-clock time. On the other hand, too large number of replicas requires a lot of disk space at grid sites to keep all these datasets.

CERN experiments have explored how to predict data popularity using machine learning and use it for the data replication strategy. ATLAS uses Artificial Neural Networks (ANN) to predict number of accesses to the dataset for the next week [2]. Access frequency represents the popularity of the dataset. One replica is then removed for unpopular datasets and new replica is created for a popular dataset once a week. 

CMS predicts dataset popularity in up-coming week using classifiers. In this study, several popularity definitions were explored [3]. Also, different classifiers were used to predict the popularity.

LHCb data management system (DMS) [4] uses a static replication strategy: a dataset has a given number of replicas during its life-time. This number can be changed by the data manager based on the knowledge of the storage occupancy and the needs for the free space. However this is not always done in an optimal way.

The first study of using machine learning for the data replication optimization in LHCb [5] was done in 2015. The approach presented in this paper is an improved and redesigned evolution of the previous one. In the new model, both long-term and short-term data popularity predictions are considered. The predictions help to detect when the number of replicas of a dataset can be increased or decreased. 

\section{General Concept}

\begin{figure}
\begin{center}
\includegraphics[width=4in]{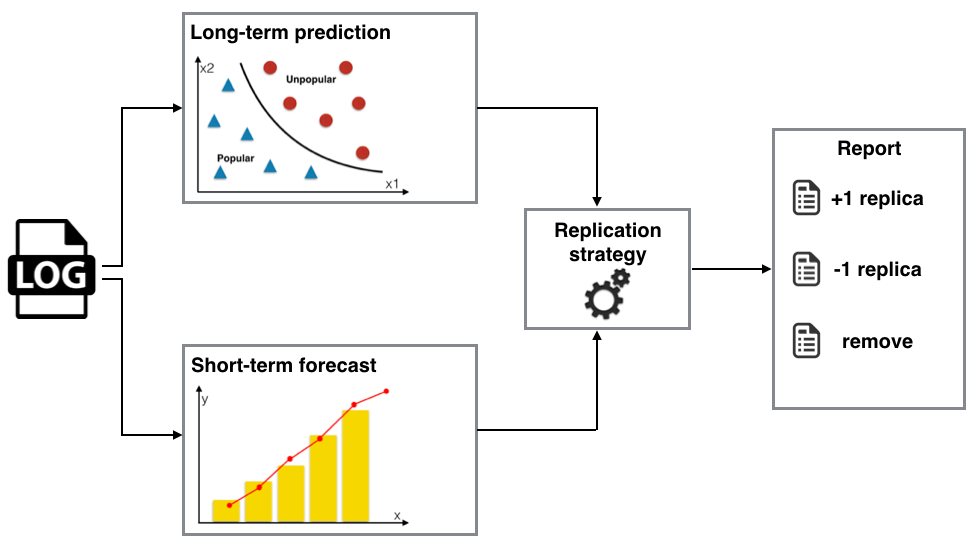}
\end{center}
\caption{\label{label}General concept of the study}
\end{figure}

LHCb has more than 15 PB of data on disk. Driven by LHCb DMS [4], each dataset is replicated on disk and also has one archival replica on tape. 

The model presented here is designed in a cache algorithms way [6]. These algorithms use specific metrics to select the most appropriate element to be removed from the cache. For example, the Least Recently Used (LRU) algorithm [6] removes the least recently used element from cache, but the Least Frequently Used (LFU) algorithm [6] removes the least frequently used one. Particular metrics used in our model will be considered in sections 3 and 5.

The study is based on the dataset access history for the last 2.5 years and its metadata. The general schema of the model is shown in Figure 1. The model provides long- and short-term predictions of the data popularity. First machine learning is used to predict the probability that a dataset will be accessed in the long-term future i.e. in the next 6 months or a year. Then a time series analysis is used to predict the number of accesses to the dataset in the short-term future: one week, one month. Then, the metrics for the replication strategy are calculated based on these predictions. This metric determines which replica or whole dataset should be removed from disk first.

\section{Long-term Prediction}
Each dataset is described by the following features: recency, reuse distance i.e. time between the last and the second last usages, time of the first access, creation time, access frequency, type, extension and size. Some of them are taken from the datasets metadata, others  are extracted from the access history. These features are used to predict the probability that the dataset will be accessed during the next 6 months.

The probability is predicted using the Random Forest Classifier. The classifier training includes two steps. In the first step, the dataset feature values are estimated using the first 1.5 out of the last 2.5 years of the access history. In the second step, the next 6 months of the history are used to label the datasets. A dataset will be labeled as popular if it is accessed during these 6 months,  and is labeled as unpopular otherwise. Then, the classifier is trained to separate the popular datasets from the unpopular ones using these features.

The classifier quality is evaluated in the following way. New feature values of the datasets are estimated using the first 2 out of the last 2.5 years of the access history. According to these values, the classifier predicts the probability that a dataset will be popular. Then, the dataset true labels are extracted from the next 6 months of  data history. Obtained predictions and the true labels are used to measure the classifier quality.

Finally, the classifier is retrained using the first 2 years of the history and uses the entire history to predict the probabilities for the next 6 months in the (unknown) future in the way described above.

Each dataset feature has a different importance for this two class separation. The Random Forest Classifier allows to evaluate this contribution. Figure 2 shows feature importances as estimated by the classifier. The figure shows that recency is the most important feature in separating popular datasets from the unpopular ones. The LRU algorithm is therefore chosen as a reference for comparison with the classifier. 

\begin{figure}
\begin{center}
\begin{minipage}[b]{3in}
\includegraphics[width=3in]{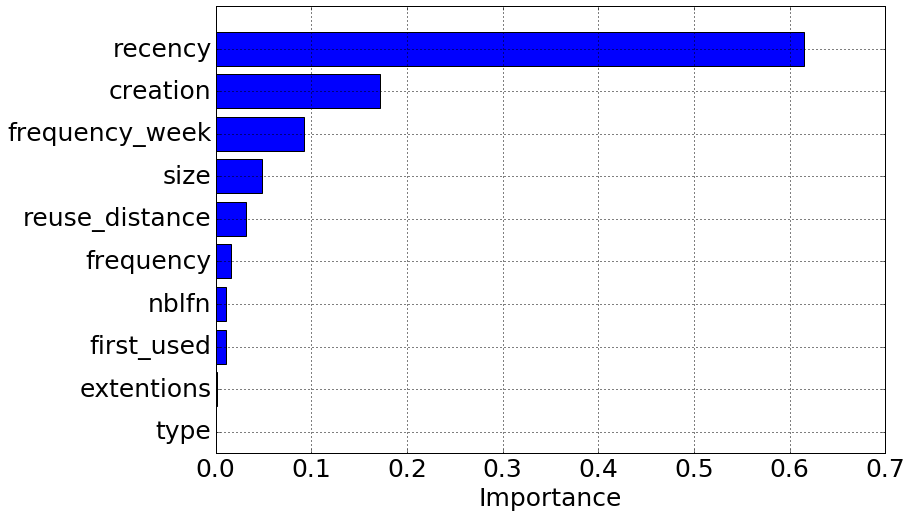}
\caption{\label{label}The dataset features importances}
\vspace{1pc}
\end{minipage}
\hspace{.2in}
\begin{minipage}[b]{3in}
\includegraphics[width=3in]{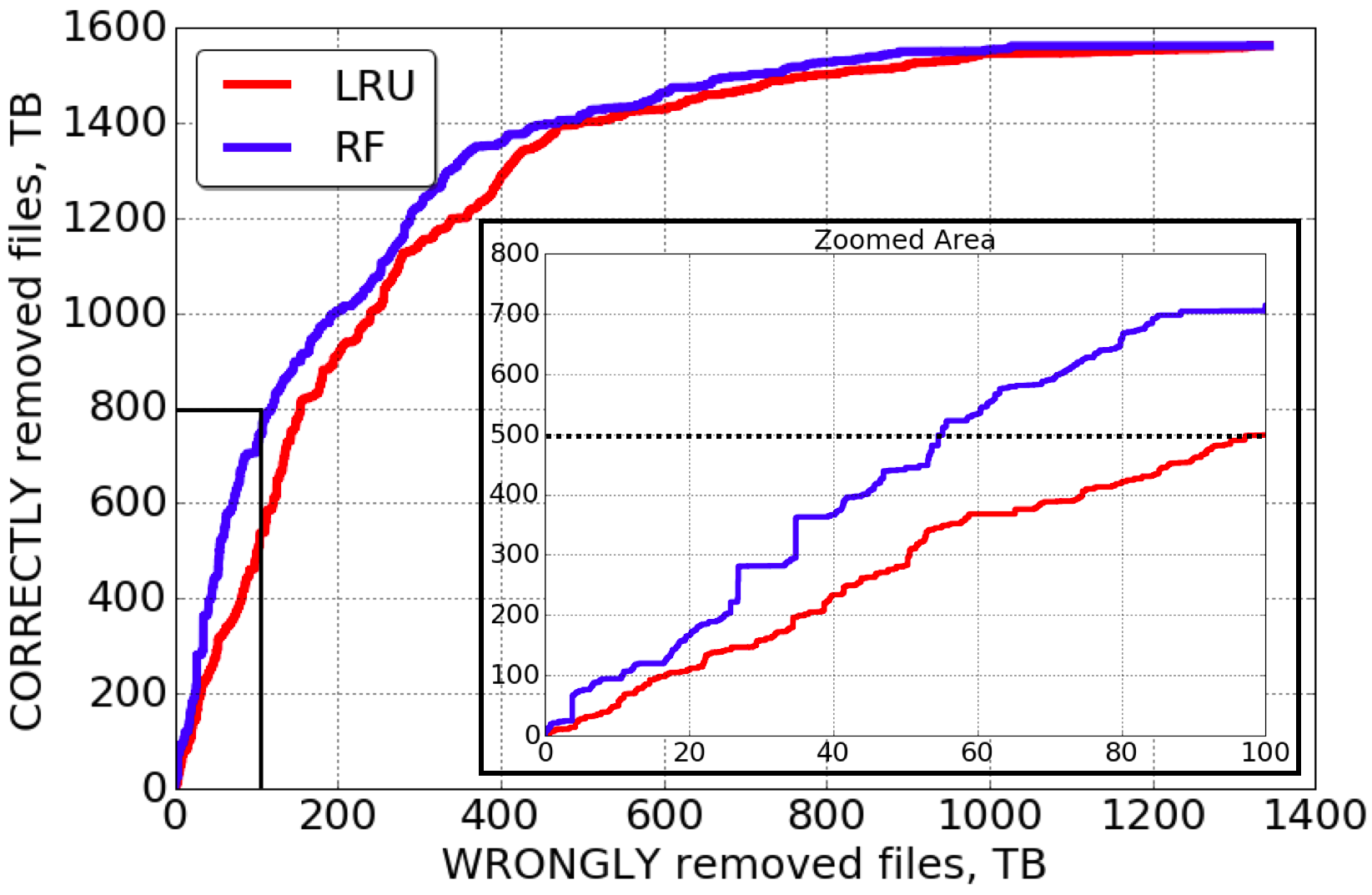}
\caption{\label{label}The dependence of the amount of the saved disk space from the mistakes.}
\end{minipage}
\end{center}
\end{figure}

The predicted probability can be used as a specific metric to save disk space. When it is needed to save some space, a dataset with the smallest access probability will be removed first. A mistake may occur, when a dataset is removed but is accessed later. In this case, this dataset should be restored to disk from its tape archive. The quality of the approach is shown in Figure 3. The figure demonstrates that the classifier allows to save the same amount of the disk space with up to 2 times smaller level of mistakes compared to the LRU method.

\section{Short-term Forecast}
In this part of the study, the time series analysis is used to predict the number of accesses to the dataset  during the coming 4 weeks. Brown's simple exponential smoothing model is chosen for that. The model is defined as:

\begin{equation}
\hat{y}_{t+1} = \hat{y}_t + \alpha (y_{t} - \hat{y}_{t})
\end{equation}

\begin{equation}
\alpha = \textit{argmin} \sum_{t} (\hat{y}_t - y_{t})^{2}, \quad \alpha \in [0, 1]
\end{equation}

\begin{equation}
\hat{y}_{0} = \frac{1}{n} \sum_{t=0}^{n} y_{t}
\end{equation}

where:
\begin{itemize}[label=]
    \item $\hat{y}_{t+1}$: is the predicted number of accesses at time-step $t+1$
    \item $y_{t}$: is the number of accesses at time-step $t$
    \item $\hat{y}_{t}$: is the predicted number of accesses at time-step $t$
    \item $\alpha$: is a smoothing coefficient
    \item $\hat{y}_{0}$: is the initial value of the prediction
\end{itemize}

The model has two extreme cases. The first one is an average model which corresponds to $\alpha = 0$. In this case, the prediction at the next time-step is the average of the values at all previous time-steps. When $\alpha = 1$ the model degenerates into a static one: the predicted value at the next time-step equals to the value at the current time-step. The actual Brown's model is somewhere in between the average and the static extremes.

\begin{figure}
\begin{center}
\begin{minipage}[b]{3in}
\includegraphics[width=3in]{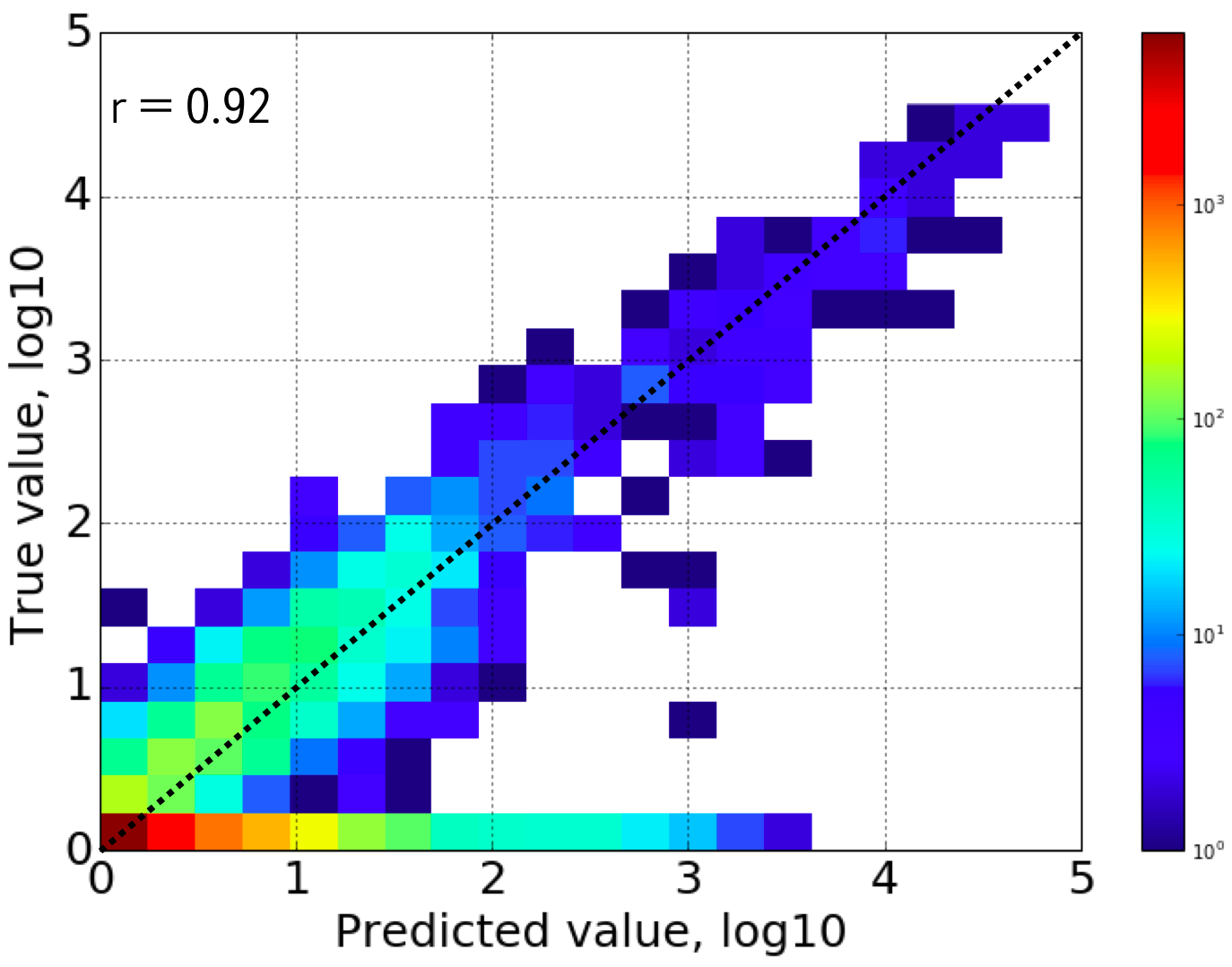}
\caption{\label{label}Short-term forecast quality}
\vspace{2pc}
\end{minipage}
\hspace{.2in}
\begin{minipage}[b]{3in}
\includegraphics[width=3in]{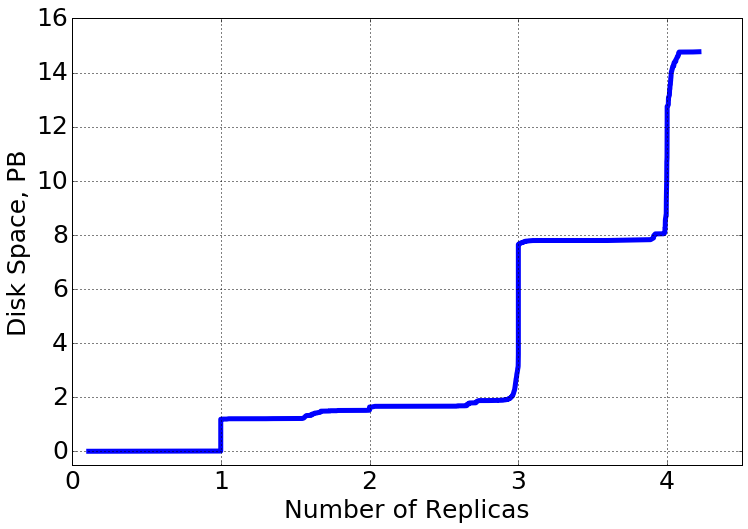}
\caption{\label{label}Cumulative distribution function of occupied disk space in LHCb from the dataset number of replicas}
\end{minipage}
\end{center}
\end{figure}

The model has only one adjustable parameter. As a result it works well with short and sparse time series. Moreover, Brown's model is simple to interpret. The figure 4 represents the quality of the model on used data. Predictions and corresponding true values demonstrate a high correlation of 0.92. The correlation for the static model is 0.86.

\section{Replication Strategy}
Based on the results of the short-term forecast the following metric is computed for each dataset:

\begin{equation}
M = \frac{\hat{y}_{t_{curr} + 1}}{n_{replicas}}
\end{equation}

where:
\begin{itemize}[label=]
    \item $\hat{y}_{t_{curr} + 1}$: is the predicted number of accesses at the next time-step
    \item $n_{replicas}$: is a dataset number of replicas
\end{itemize}

This metric is used to adjust the number of dataset replicas. A replica for a dataset with the minimum metric value is removed first when it is required to free some disk space. The very last replica of the datasets is not removed and stay on disk. After each deletion the $M$ value is recalculated and the procedure is repeated. 
When there is some available free disk space, a new replica for the dataset with the maximum metric value is added first. Then, the metric value is recalculated and the procedure is repeated.

Figure 5 represents the cumulative distribution of occupied disk space in LHCb for different number of replicas. The function shows that the largest part of the space actually corresponds to datasets with 3 and 4 replicas. Thus, reducing number of replicas for them does allow saving petabytes of disk space when it is needed.

Generally, the strategy described above allows to use the disk space more optimally. However once or twice per year it makes sense to remove also the very last replicas for those datasets which are the most unpopular based on the long-term predicted probabilities as described in Section 3. This saves the disk space occupied by the datasets which most likely will never be accessed in the future.

\section{Conclusion}

The paper presents the approach of how machine learning and time series analysis can be used in data replication strategy. Forecast of the number of accesses of datasets allows to optimize the number of replicas. The prediction of the probability for  the dataset to be used in long-term future helps to remove the least popular data from disks. These methods outperform conventional ones such as static model and LRU, and helps to use disk space more optimally with lower risk of mistakes and data loss.  

\section{Acknowledgments}

The authors are grateful to Denis Derkach and Fedor Ratnikov for their helpful comments during the work on this study and help with preparing this paper.


\section*{References}
\begin{thebibliography}{9}
\bibitem{iopartnum} Stagni F, etc. 2012 LHCbDirac: distributed computing in LHCb {\it J. Phys.: Conf. Series} {\bf 396} 032104 (IOP Publishing)
\bibitem{iopartnum} Beermann T, Stewart G A and Maettig P 2014 {\it The International Symposium on Grids and Clouds (ISGC) 2014: A Popularity-Based Prediction and Data Redistribution Tool for ATLAS Distributed Data Management}
\bibitem{iopartnum}  Kuznetsov V, Li T, Giommi L, Bonacorsi D and Wildish T 2016 Predicting dataset popularity for the CMS experiment {\it J. Phys.: Conf. Series} {\bf 762} 012048 (IOP Publishing)
\bibitem{iopartnum} Baud J P, Charpentier P, Ciba K, Graciani E, Lanciotti E, Màthè1 Z, Remenska D and Santana R 2012 The LHCb Data Management System {\it J. Phys.: Conf. Series} {\bf 396} 032023
\bibitem{iopartnum} Hushchyn M, Charpentier P and Ustyuzhanin A 2015 Disk storage management for LHCb based on Data Popularity estimator {\it J. Phys.: Conf. Series} {\bf 664} 042026 (IOP Publishing)
\bibitem{iopartnum} Saemundsson T 2012 An experimental comparison of cache algorithms \url{http://en.ru.is/media/skjol-td/cache_comparison.pdf}
\end{thebibliography}

\end{document}